\documentclass[twocolumn]{aastex631}
\usepackage{amsmath,amstext,chngcntr}
\usepackage[T1]{fontenc}
\usepackage{hyperref}
\usepackage{ae,aecompl}
\usepackage[utf8]{inputenc}
\usepackage{apjfonts} 
\usepackage[figure,figure*]{hypcap}
\usepackage{enumitem}
\usepackage{bm}
\usepackage{pifont}
\usepackage{slashed}
\input{hyperlink-year-only-natbib-patch}

\newcommand{\nn}{\textbf{n\_neighbours}}
\newcommand{\md}{\textbf{min\_dist}}
\newcommand{\nc}{\textbf{n\_componenets}}
\newcommand{\met}{\textbf{metric}}
\newcommand{\Mpch}{\ensuremath{\rm{Mpc}~h^{-1}}}

\newcommand{\pch}{\ensuremath{\rm{pc}~h^{-1}}}
\newcommand{\Smass}{\ensuremath{M_{\odot}}}
\newcommand{\Smassh}{\ensuremath{M_{\odot} h^{-1}}}

\newcommand{\subrvir}{\ensuremath{R_{\rm vir,sub}}}

\begin{document}

\title{ A study of the dynamical structures in a Dark Matter Halo using UMAP}  
\shortauthors{Narayan R. $\&$ Adhikari}
\shorttitle{Halo structures using UMAP}

\author[0009-0009-8338-3388]{Soorya Narayan R.}
\affiliation{Department of Physics, Indian Institute of Science Education and Research, Dr. Homi Bhabha Road, Pune, 411008, India}

\author[0000-0002-0298-4432]{Susmita Adhikari}
\affiliation{Department of Physics, Indian Institute of Science Education and Research, Dr. Homi Bhabha Road, Pune, 411008, India}

\begin{abstract}

We use a dimension reduction algorithm, Uniform Manifold Approximation and Projection (UMAP), to study dynamical structures inside a dark matter halo. We use a zoom–in simulation of a Milky Way mass dark matter halo, and apply UMAP on the 6 dimensional phase space in the dark matter field at $z=0$. We find that particles in the field are mapped to distinct clusters in the lower dimensional space in a way that is closely related to their accretion history. The largest cluster in UMAP space does not contain the entire mass of the Milky Way virial region and neatly separates the older halo from the recently accreted matter. Particles within this cluster, which only comprise $\sim 70\%$ of the Milky Way particles, have had several pericenter passages and are, therefore, likely to be phase mixed, becoming dynamically uniform. The infall region and recently accreted particle and substructure, even up to splashback, form distinct components in the lower dimensional space; additionally, higher angular momentum particles also take longer times to mix. Our work shows that the current state of the Milky Way halo retains historical information, particularly about the recent accretion history, and even a relatively old structure is not dynamically uniform. We also explore UMAP as a pre-processing step to find coherent subhalos in dark matter simulations.

\end{abstract}


\section{Introduction} \label{sec:intro}

Dark matter halos form the potential wells within which galaxies form and evolve in the universe \citep{COORAY_2002, 1978MNRAS.183..341W}. Halos form from the gravitational collapse of initial density peaks \citep{1984Natur.311..517B,1985ApJ...292..371D, Bahcall_1999} in the dark matter field and form bound, self-gravitating objects that are thought to relax through violent relaxation and phase mixing\citep{binney_tremaine_2008}. As the dark matter field in the early universe is drawn from a spectrum of perturbations that follow a Gaussian random field \citep{1986ApJ...304...15B}, the full non-linear evolution of a dark matter halo, coevolving in this field, is a dynamic and complicated process. In the cold dark matter (CDM) paradigm, structure formation proceeds hierarchically, i.e. smaller mass objects form first and merge to form larger halos. This process leaves behind, within a single dark matter halo, a large range of smaller substructures called subhalos, which tidally disrupt and deposit their mass throughout the host's internal volume. Additionally, large-scale tidal fields affect the shapes of the local density distribution \citep{Hayashi_2003, Pe_arrubia_2010, Green_2019, Errani_2021}, making isolated evolution of halos significantly different from evolution in the cosmological environment. Eventually, however, a universal profile appears to emerge, and this is well-fitted by what is known as the Navarro Frenk $\&$ White (NFW) profile \citep{Navarro:1996gj}

While NFW is a fairly good fit to the mean across a wide range of structures, there is significant scatter from the universal profile both in the inner and outer regions for halos of different masses expected to originate from unique accretion histories \citep{ FG84, LD10, Diemer:2014xya}. Typically, halos are defined as regions inside what is known as the halo virial radius. This radius corresponds to an overdensity at which a spherical top-hat perturbation is expected to reach virialization \citep{1969PThPh..42....9T,1972ApJ...176....1G}. This boundary, while physically motivated, is somewhat arbitrary. More recently, the splashback radius \citep{Diemer:2014xya, Adhikari:2014lna, More:2015ufa, Shi:2016lwp}, which is the phase space boundary formed by the apocenter of the most recently accreted, has been proposed as a more physical definition of a dynamical boundary of the halo. This radius is also closely related to the radius of lowest radial velocity \citep{Cuesta:2007it} and is similar to such physical regions encompassing the multi-streaming region of the halo \citep{Aung:2020czp, Garcia:2022zsz, Zhou:2023uhb}. For individual halos that are not necessarily spherical, the splashback shell still forms a boundary between multi-streaming and the infall region \citep{Mansfield:2016bxx, 2017ApJS..231....5D, 2020ApJS..251...17D}.  However, it is worth asking whether the structure, even within such a boundary, is dynamically uniform and how long subhalos and streams remain distinct structures within such a boundary in their host. 

\begin{figure*}
\includegraphics[width=\textwidth]{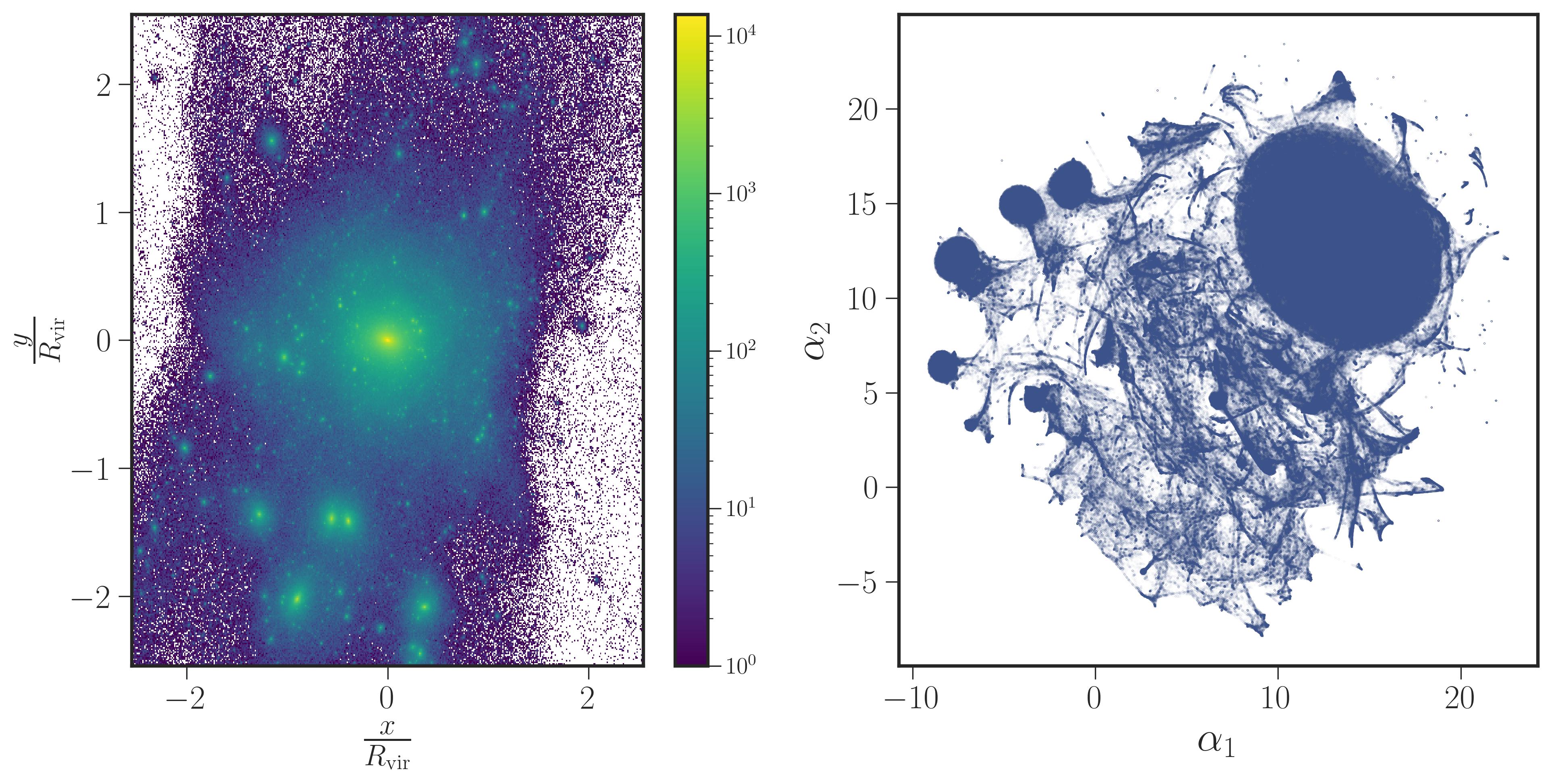}
\caption{\textbf{Left :} Color density map of a 1 Mpc $h^{-1}$ region around the halo scaled by the virial radius.  \textbf{Right :} Lower dimensional representation of the 6D phase space constructued using the UMAP algorithm. UMAP creates a mapping to optimally separate dynamically uniform clusters in higher dimensional space into distinct clusters in a projected space. We use $\nn=30$ for this projection (see text in Section \ref{sec:UMAP}).}
\label{fig:Prelim}
\end{figure*}

CDM N-body simulations are a powerful way to study the full non-linear evolution of dark matter density field starting from cosmological initial conditions in a $\Lambda$CDM universe \citep{1983MNRAS.204..891K, 1985ApJS...57..241E} (see also \cite{Bagla:2004au, Angulo:2021kes} for a review). Zoom--in simulations are employed to study single objects in high resolution, where the large-scale models of the gravitational field are drawn from a lower-resolution, larger scale cosmological volume. These simulations track the full history of individual dark matter simulation particles from initial conditions to their final state at a given redshift. The wealth of data provided by these methods is invaluable to explore the full complex structure of dark matter halos and has formed the backbone of modeling non-linear structure formation and galaxy--halo connection \citep{Wechsler:2018pic}. Typically, halo definitions based on virial radius are employed to isolate overdense structures and identify them as halos and subhalos to study their inter-relations both among themselves and with the large-scale structure \citep{ 2001MNRAS.328..726S, 1999ApJ...516..530K, 2009ApJS..182..608K, Behroozi_2012A}.

In this paper, we explore some of the questions about the internal structure of a halo and the dark matter field around it using alternate, clustering and machine learning techniques. We use the full 6D phase space information of these particles to find dynamically coherent structures around a dark matter halo. We use a technique called Uniform Manifold Approximation and Projection for Dimension Reduction (UMAP)\footnote{The UMAP package website is \href{https://umap-learn.readthedocs.io/en/latest/index.html}{UMAP docs}} \citep{mcinnes2020umapuniformmanifoldapproximation}, which is a graph-based clustering and machine learning algorithm, to find dynamically distinct groups that remain agnostic to any pre-definitions of halo boundaries. 
UMAP is a non-linear dimension reduction algorithm that projects high-dimensional data onto a lower-dimensional hyperplane while preserving topology. It is primarily used to visualize different groupings within data in higher dimensions in an interpretable way. We find that it is an ideal tool to study how the current structure of the halo encodes evolutionary information in the field.
This paper is organized as follows: In Section \ref{sec:data}, we introduce the N-body simulations we are using. In Section \ref{sec:UMAP}, we describe the UMAP algorithm. In Section \ref{sec:results}, we describe the interpretation of the results and the separation of structures in UMAP space. We discuss and conclude in Section \ref{sec:conclusions}.

\section{Simulations} \label{sec:data}
For this study, we use zoom--in simulations of Milky Way-mass halos from the suite of Milky Way mass halos run in \citep{Mao_2015}. The zoom--in region is selected from a $125\Mpch$ box with $1024^3$ particles with $\Omega_m = 0.286$, $\Omega_\Lambda = 0.714$, $h = 0.7$, $\sigma_8 = 0.82$ and $n_s = 0.96$. The initial conditions of the zoom--in simulations are generated using the publicly available MUSIC code \citep{2011MNRAS.415.2101H} and are matched to the cosmological box to the $1024^3$ scale. The zoom--in simulation  consists of 236 snapshots between $z=19$ and $z=0$. The mass of the highest-resolution particles in the zoom--in simulation is $2.8 \times 10^5 \Smassh$. The softening length in the highest-resolution region is $170 ~\pch$ comoving.

We use the ROCKSTAR \citep{Behroozi_2012A} and Consistent Trees \citep{Behroozi_2012B} to identify halos and make a subhalo catalog. Note that the Milky Way is also the most massive halo in this region, with a mass of $8.7 \times 10^{11}\Smass$ , a virial radius of $0.2 ~\Mpch$ and a maximum circular velocity of $171.81 {\rm km ~s}^{-1}$ at $z=0$. The virial region is defined as $\Delta_{\rm vir}=363 \bar\rho_m $ . From here on out, we refer to this halo as the Milky Way (MW). 

For this study, we primarily focus on a $1\Mpch$ comoving region around the center of the MW. MW center is at the center of the box. We extract the particles, their complete phase space information $[x_i, v_i]$, and the complete trajectory of these particles for analysis and interpretation of our results. We have a total of 4,990,295 particles in the $1~\Mpch$ box, 3,112,172 of which lie within the virial radius of the MW halo. In the next section, we describe the UMAP representation of the dynamical halo.

\section{UMAP representation of the Milky Way Halo}
\label{sec:UMAP}
\begin{figure*}
    \centering
    \includegraphics[width=1\textwidth]{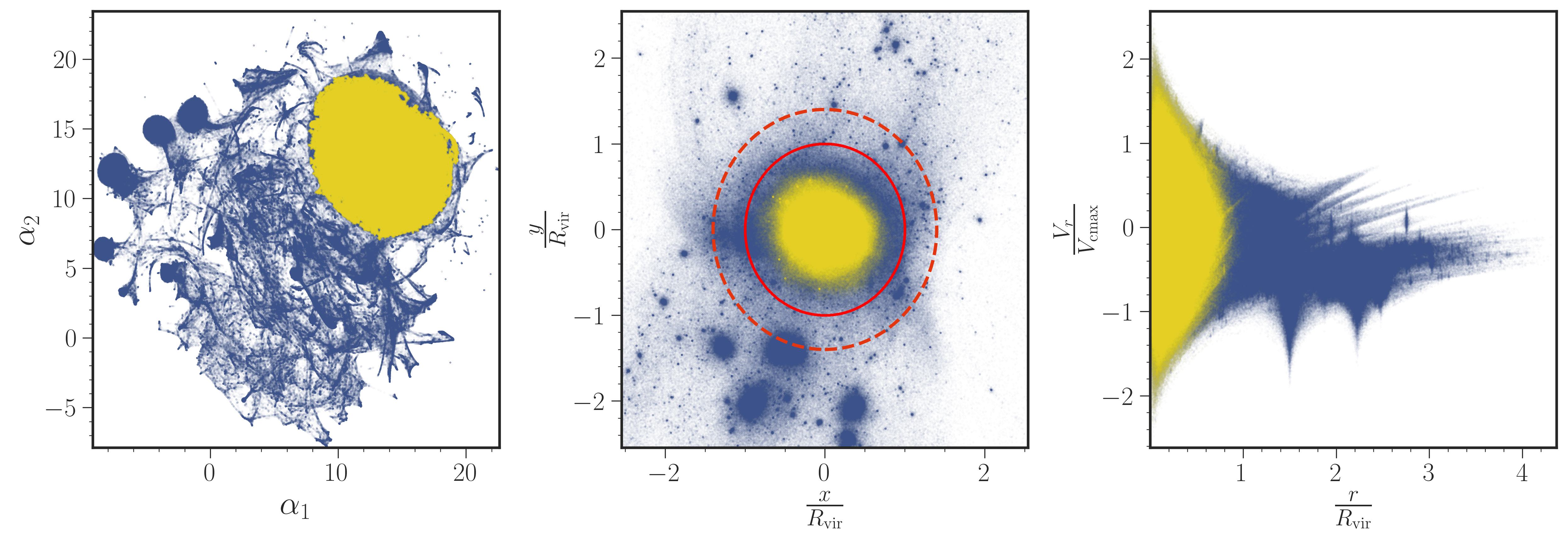}
    \caption{\textbf{Left : }The yellow points show the particles belonging to the largest cluster in UMAP space identified using DBSCAN overlaid on the full UMAP representation (blue points).
    \textbf{Center : }The same points are plotted in configuration space plotted in the $x-y$ plane against the rest of the particles in the box (deep blue). The red circle represents the estimated virial boundary of the MW. The red dashed circle represents the estimated splashback radius. \textbf{Right : }shows the phase space ($V_r-r$) plot of the same cluster (yellow) against the phase space distribution of the entire box (deep blue).}
    \label{fig:Ellipse}
\end{figure*}

The 6D phase space information of all the particles in our $1 \Mpch$ box (at $z=0$) is the input to the UMAP algorithm. UMAP is a manifold learning technique that uses Riemannian geometry and algebraic topology \citep{mcinnes2020umapuniformmanifoldapproximation}. It is often used to see patterns and interconnections between groups in higher dimensional space in a lower dimensional space. In practice, UMAP constructs a weighted graph in the higher dimensional input space where each edge is weighted by the distance between the two nodes. The metric used to calculate the distance is allowed to vary from node to node based on the local density of points. This variation allows for encoding the topology of the structures in the input space into the weighted graph (we will refer to this graph as $\mathcal{G}$). This step of constructing the input manifold produces the topological representation of the input data. Given an input map $\mathcal{G}$, the projection begins with a spectral embedding \citep{10.5555/2980539.2980616} in the low-dimensional space. 
UMAP then calculates the weighted graph of said representation, $\mathcal{H}$, and proceeds to minimize the ``fuzzy set cross entropy'' between $\mathcal{G}$ and $\mathcal{H}$, by moving the points around in the low dimensional representation. 
Fuzzy cross-entropy is an estimate of the mismatch between $\mathcal{G}$ and $\mathcal{H}$. For every edge $e$ in the set of all edges $\mathrm{E}$, if $w_h(e)$ denotes the weight in the higher dimensional representation and $w_l(e)$ denotes the weight in the lower dimensional representation, then the cross entropy is given by,
\begin{equation}
    \mathcal{S}=\sum_{e\in\mathrm{E}}w_h(e) \thickspace \log\left[\frac{w_h(e)}{w_l(e)}\right] + (1 - w_h(e)) \thickspace \log\left[\frac{1-w_h(e)}{1-w_l(e)}\right] \label{cross_entropy}
\end{equation} 
The latter step of UMAP is a learning process. Along with the low-dimensional representation, UMAP also outputs the learned mapping function, $\mathcal{\Phi}$, between the input and output spaces that produce said low-dimensional representation with the lowest fuzzy cross-entropy.

There are four main parameters for UMAP that decide the low-dimensional projection. $\nn$ decides the scale at which UMAP probes structures in the input space. It dictates the size of the local neighborhood where UMAP will attempt to learn the manifold structure. This neighborhood decides how the local metric is scaled for each data point, resulting in a different estimation of $\mathcal{G}$ for the input data for each value of $\nn$. Smaller values of $\nn$ identify very local structures, while larger values result in more global patterns. $\md$ decides how tightly the data points can be packed in the low-dimensional space by modifying $\mathcal{H}$. Unlike $\nn$, $\md$ does not affect $\mathcal{G}$.
$\nc$ is the target dimension. $\met$ decides what metric is used to calculate distances for $\mathcal{G}$ and $\mathcal{H}$. A more detailed description of UMAP's algorithm is given in Appendix \ref{app:umap}.

Note that the position values we provide are taken with respect to the center of the MW and scaled with the virial radius of the MW. The velocity information is also taken with respect to the velocity of the MW center and scaled by the maximum circular velocity of the MW at $z=0$. Normalization or scaling is ubiquitous in machine learning as it ensures equal contribution from all input features, preventing features with large magnitudes from dominating the predictions. It also enables faster convergence for any gradient-descent-based algorithms (which is part of the optimization process).

For our fiducial case, after some trial and error, we chose an $\nn$ value of 30. This value corresponds to the peak of the halo mass distribution (from ROCKSTAR) in the box. We use Euclidean $\met$, set $\nc$ to 2, i.e., we choose a two-dimensional representation for interpretability and $\md$ to 0. Setting $\md$ to 0 produces compact clusters in the output space. This choice also makes cluster identification easier in UMAP space (Section \ref{subsec:dynamical_structures}). The most important parameter for our purposes is $\nn$, Fig. \ref{fig:appendix_umap_nn} shows how the structures vary as we change this parameter. 

Figure \ref{fig:Prelim} shows the density color map of the Milky and the corresponding UMAP representation derived from the 6D space of the selected particles. Each particle in real space has a corresponding particle in the output UMAP space. Since it is a non-linear dimension reduction algorithm in which the local metric varies from point to point, the two variables, $\alpha_1$ and $\alpha_2$, by themselves, have no physical meaning. Therefore, the image on the right is simply an arrangement to preserve the topology of the 6D phase space distribution of the particles on the left, and each cluster of particles corresponds to a different group of objects in the higher dimensional space. 

The primary features that appear in the UMAP representation are as follows. Firstly, we observe that the field is dominated by a relatively smooth, single large elliptical structure on the top-right. Multiple, relatively smaller clusters are placed separated from the larger ellipse and are connected by elongated filamentary structures. The large structure naturally corresponds to the majority of the particles in the Milky Way, and the few other large objects can be identified as some of the major infalling structures in the vicinity of the central halo. Each large cluster seems to have a degree of self-similarity in the density and arrangement of particles around them in the UMAP space. In the next section, we will delve into understanding the features of the halo field in UMAP space in more detail.

\begin{figure*}
    \centering
    \includegraphics[width=\textwidth]{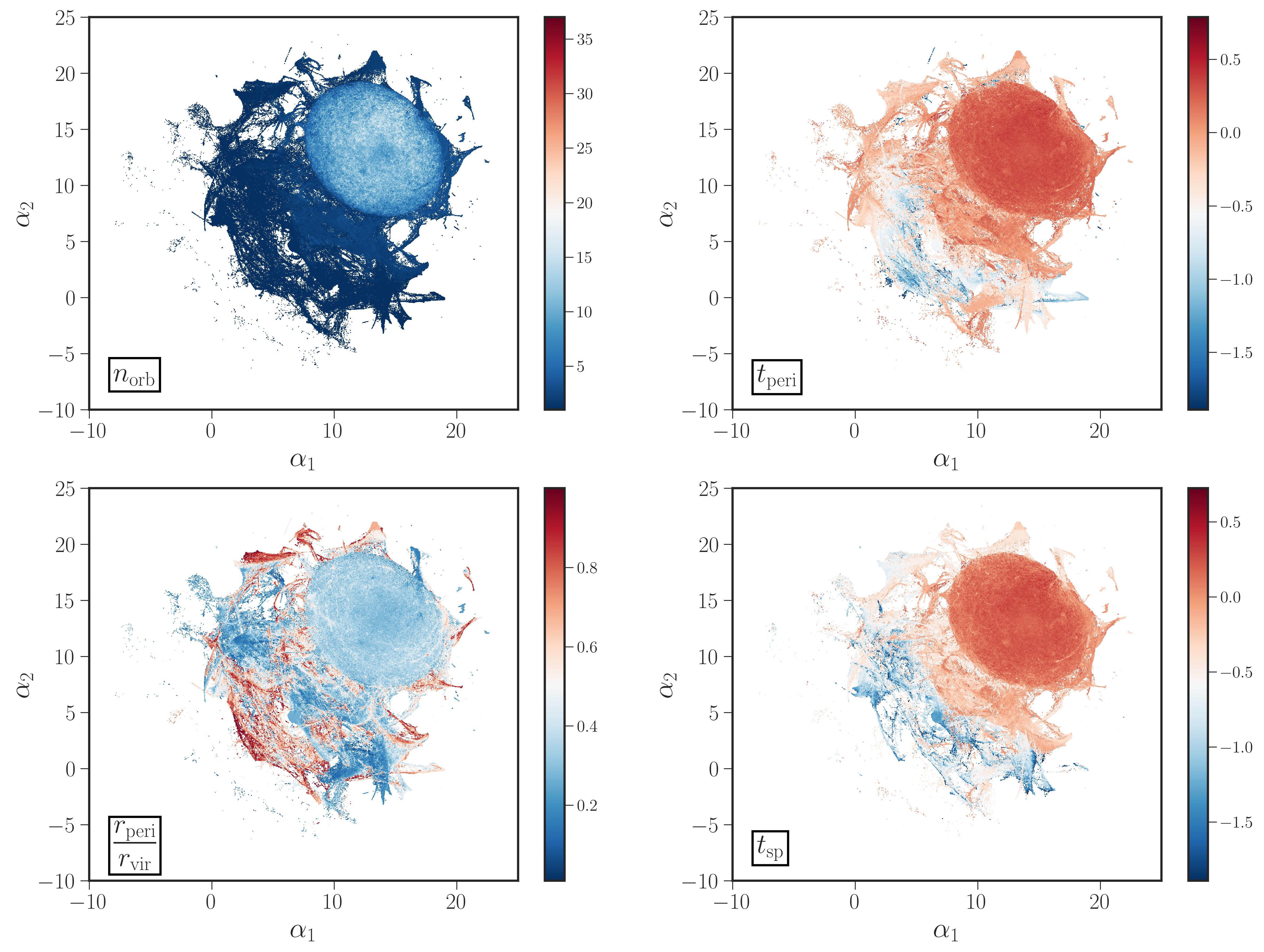}
    \caption{Color Density plot in UMAP space where the colors represent the mean value of various properties related to the history of the particle trajectory. \textbf{Top left : }shows the number of orbits completed by the particles by $z=0$. 
    \textbf{Top right : }the colors correspond to the mean logarithmic redshift of the first pericenter passage of all particles in that grid point. Lower, darker values values correspond to more recent times.
    \textbf{Bottom left : }shows the particles' pericenter distance in units of the viral radius of the MW at the time of the pericenter passage.
    \textbf{Bottom right : }  shows the mean redshift at which particles reach splashback in that bin. Lower, darker values values correspond to more recent times.} \label{fig:pericenter}
\end{figure*}

\section{Interpretation} 
\label{sec:results}
\subsection{Separation of dynamical structures}
\label{subsec:dynamical_structures}

For our purposes, it is insightful to understand the physical origin of the different structures that arise in the UMAP representation space of the box. Clustered regions are isolated in this lower dimensional space and studied further for interpretability.

To begin with, we start by studying the largest structure in the UMAP space. Naturally, we expect that this object is associated with the most gravitationally dominant body in the box, which in our case is the central Milky Way dark matter halo. UMAP gives a one-to-one mapping of each particle. We identify the particles belonging to the largest structure in UMAP space and map it back to the configuration space and the $V_r-r$ space\footnote{We will refer to this as the radial phase space from here on.}. This is shown in Figure \ref{fig:Ellipse}. To delineate groups, we use a density-based clustering algorithm called the Density-Based Spatial Clustering of Applications with Noise (DBSCAN). It works based on the notion of high-density groups separated by low-density noise regions. 
This is an algorithm similar to FOF that associates particles within some pre-defined linking length into a single structure. We partly use trial and error to fix the linking length to extract the largest elliptical structure in the projected space.

With the choice of parameters used to pick out the large ellipse, DBSCAN classified $\sim 10\%$ of the particles as noise. Since the outer contour of the largest cluster is approximately an ellipse, we will refer to the points within this cluster as \textit{particles within the ellipse} or simply \textit{the ellipse}, and the rest of the points as \textit{particles outside the ellipse}. In Figure \ref{fig:Ellipse}, left panel, the points belonging to this ellipse are plotted in yellow and embedded on top of the full UMAP space. The central panel shows the corresponding location of the particles (yellow) in the $x-y$ plane of the simulation box. The red circle corresponds to the ``virial radius'' of the MW. We note that the ellipse is confined to the inner region of the halo, i.e. it does not encompass the full virial region of the halo; this implies that even within the erstwhile defined halo radius, not all particles belong to the same UMAP group.  The right panel shows the particles of the ellipse (yellow) in radial phase space against the rest of the particles. This space is particularly intuitive as it clearly shows the infall stream and multi-streaming regions of the halo. Particles with negative velocities are moving towards the center, and those with positive velocities are moving away. The conical region comprising of multistreaming particles is typically the boundary of orbiting particles, its vertex alternatively defines the size of the halo often referred to as the splashback radius, $R_{\rm sp}$. We note that yellow points do not extend all the way out to $R_{\rm vir}$ or $R_{\rm sp}$. 

A magnified version of the radial phase space is shown in the Appendix Figure \ref{fig:appendix_ellipse_logspace}. The particles in the ellipse form largely the smooth component of the halo confined to the inner region but extending up to nearly $0.9 R_{\rm vir}$. There are some stream-like structures that extent to larger radii, these are primarily tidally disrupted streams from recently infalling substructure that are turning around. We note that none of the infalling subhalos, despite being part overlapping in both radius and velocity in the configuration space and radial phase space, are incorporated within the ellipse. In fact, as we will point out in the next section, the UMAP algorithm largely places particles into distinct structures in a way that is strongly correlated with their orbital history.

\begin{figure*}
    \centering
    \includegraphics[width=0.9\textwidth]{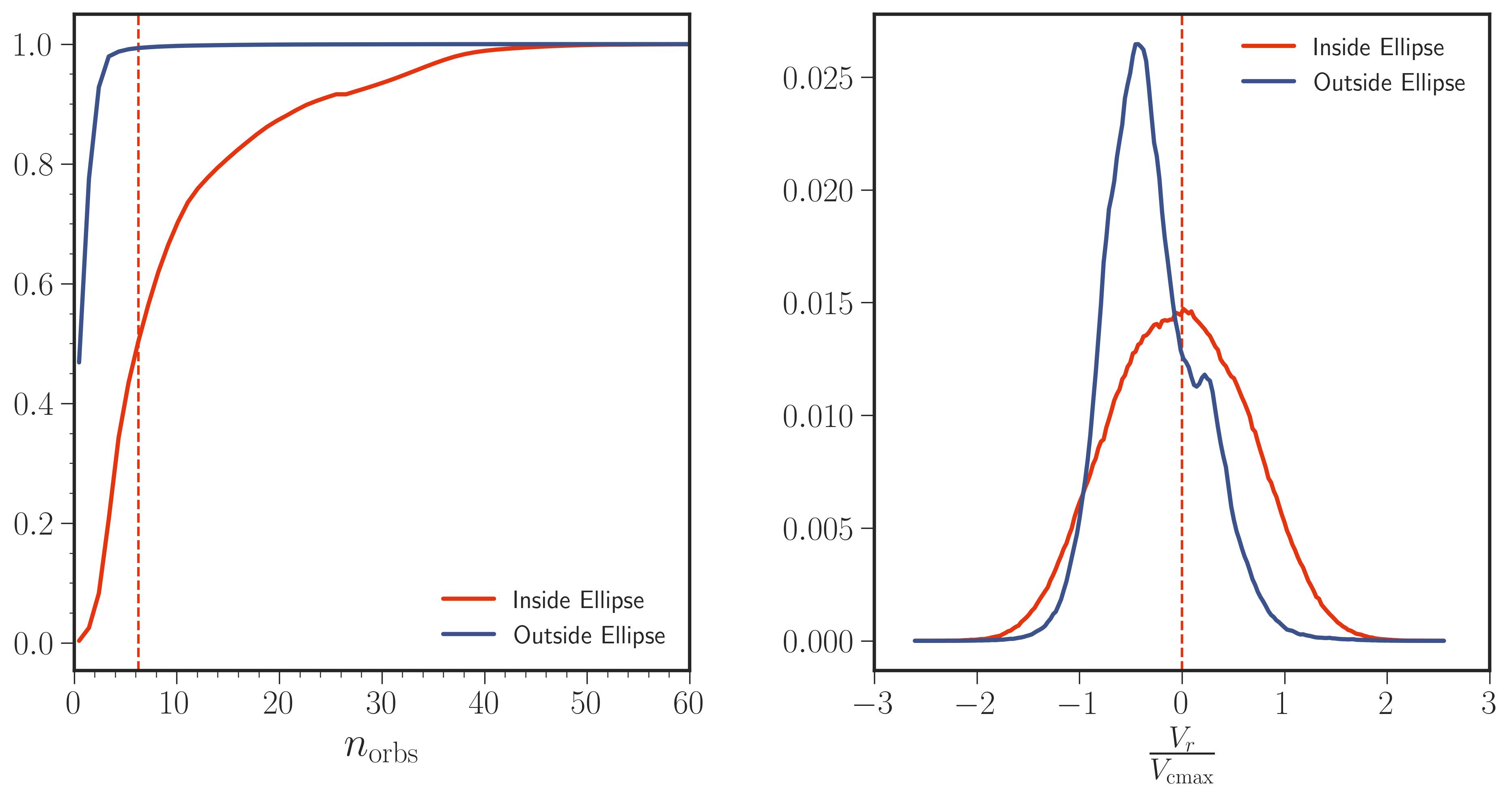}
    \caption{\textbf{Left : }shows the cumulative distribution of the number of orbits the particles, inside and outside the ellipse, have had by $z=0$. The vertical line indicates the point where the red curve measures $0.5$.
    \textbf{Right : }shows the distribution of the radial velocity of the ellipse and the rest of the particles in the box. The vertical line marks $V_r = 0$.} \label{fig:1d-distributions}
\end{figure*}

\begin{figure*}
    \centering
    \includegraphics[width=\textwidth]{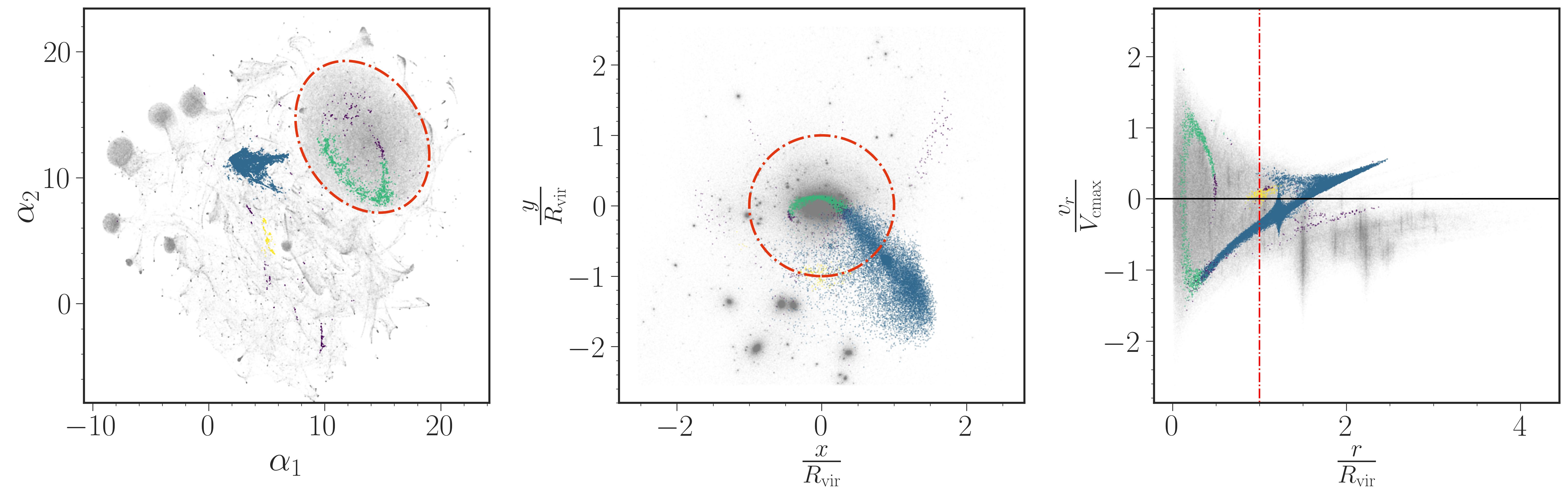}
    \includegraphics[width=\textwidth]{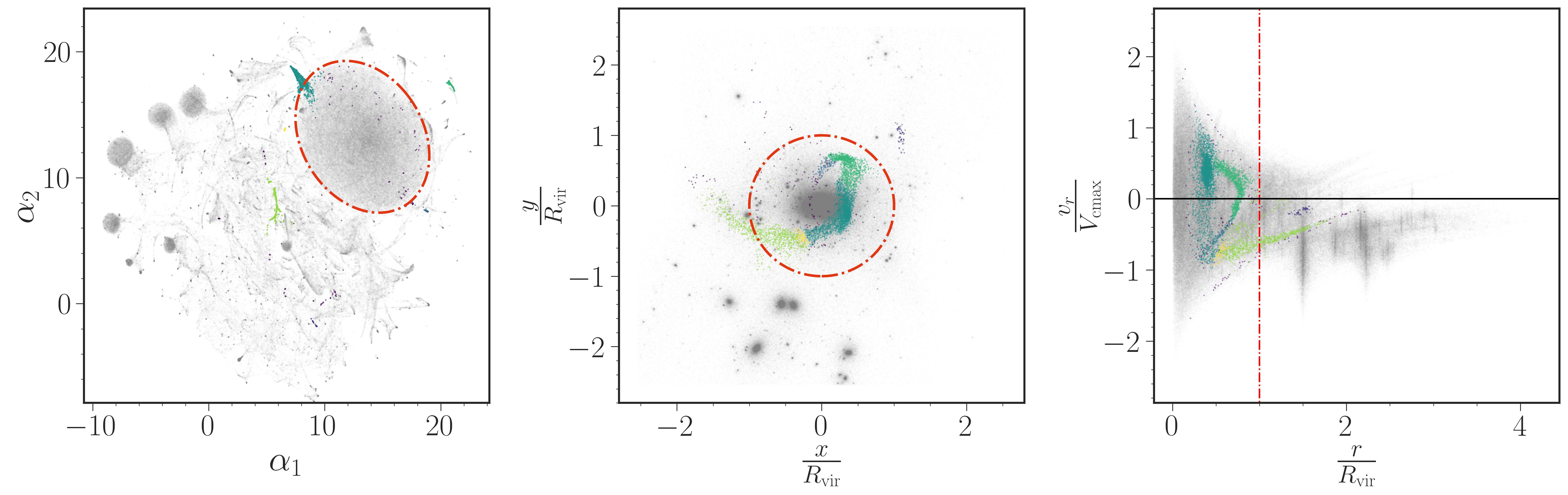}
    \caption{Subhalos in UMAP space. The red dashed curve in the left column shows the approximate boundary of the ellipse. The red dashed circle in the middle column shows the virial radius of the MW. The red dashed line in the right column represents the virial boundary of the MW, and the black horizontal line represents $V_r = 0$. The colored points correspond to subhalo halo particles. The subhalos were identified using ROCKSTAR at $z=0$. The particles belonging to the subhalo are all particles that are within towards $1.5\subrvir$ before infall (close the $M_{peak}$). The top row shows a subhalo of mass $6.4 \times 10^9 \Smass$, and the bottom row shows a subhalo of mass $1.9 \times 10^9 \Smass$.
    The grey background in each panel shows particles from the entire box downsampled by a factor of 10 for reference. The subhalo particles are clustered in UMAP space using DBSCAN to separate distinct parts of the same object. The left panel shows the subhalo in UMAP space, the middle panel shows the real space $x-y$ plane, and the leftmost panel shows the radial phase space. } \label{fig:subhalos_umap}
\end{figure*}

\subsection{UMAP cluster and orbital properties}
Here, we attempt to summarize our findings on the relation between the particle orbital properties and the UMAP clusters in projected space. We extract the positions of the particles over the full available range of snapshots and track the path of every particle in the selected region with respect to the Milky Way center. We evaluate and record the time  $(t_{\rm inf})$ when each particle enters the virial radius of the Milky Way, the location $(r_{\rm peri})$ and time $(t_{\rm peri})$ when the particle reaches pericenter, and the location $(r_{\rm sp})$ and time $(t_{\rm sp})$ when the particle reaches its first apocenter after infall or ``splashback''. We also record the number of pericenter crossings each particle has had, $n_{\rm orb}$.

Figure \ref{fig:pericenter} shows 2D heat maps of UMAP space, colored by the mean values of various relevant particle properties at that location. These maps contain particles that have had at least one pericenter crossing by $z=0$ (infalling particles are therefore left out). We find that the map that shows the colors weighted by $n_{\rm orb}$ (top left) shows the most dramatic separation in UMAP space. Here, the color represents the mean number of pericenter passages that particles in that bin have completed. Particles within the ellipse have clearly been inside the halo longer and through multiple orbits. Those outside the ellipse mostly have had fewer than two orbits. 
The top right panel shows the logarithmic redshift of pericenter crossing. Higher values indicate earlier times, while lower values indicate more recent events. Particles in the ellipse have very early pericenter times relative to the rest of the particles. Regions further from the ellipse comprise particles that have crossed their pericenter more recently. The most recently accreted particles are placed furthest from the ellipse. The bottom left panel shows the ratio of the pericenter distance to the virial radius of the halo at $t_{\rm peri}$. We note that all particles with large pericenters, i.e., higher angular momentum in their orbits, are placed outside the largest ellipse (redder colors). Comparing the top-right and bottom-left panels, there is also a visible correlation between the time of pericenter crossing and the distance of pericenter crossing, showing that more recently accreted particles tend to have higher angular momenta.

The bottom right panel shows the redshift ($\log(z)$) of the first apocenter crossing after infall, i.e., the redshift of splashback.  This typically forms the boundary of the multistreaming region in the radial phase space of dark matter particles. As expected, this map is highly correlated with the top right panel showing the $\log(z)$ values for pericenter crossing. The gradient going from the ellipse center to the outside is much more pronounced here. The particles within the ellipse form the older component of the halo and are most separated from the particles that are reaching splashback at $z=0$. The recent splashback particles also show a large amount of smaller substructure and stream-like extensions, showing that particles here are still distinct from the main smooth halo dynamically. The splashback and, more generally, the recently accreted particles appear to be more identical to the infalling structure than the smooth elliptical structure that forms the largest clustered component in the UMAP space.
Summarizing the main takeaway from these comparisons is that the separation in UMAP space is deeply linked to the relaxation process in halos through phase mixing and the equipartition of energy. As particles orbit multiple times within the potential and their phase space trajectories get wrapped around, two particles that are close in phase space are originally drawn from vastly different regions of the initial phase space. It has been known that cold dark matter halos relax through this procedure \citep{binney_tremaine_2008}. The older phase-mixed structure appears to form a distinct cluster in this space, retaining a distinct identity from the more recently accreted halo. The inner ellipse also includes subhalos that have been inside the host for multiple orbits.

Figure \ref{fig:1d-distributions}, shows the cumulative distribution of no. of orbits of particles within the ellipse compared to those without. The peak of the distribution within the ellipse is at $n_{\rm orb}=4$, and the median is at an even higher value. Outside the ellipse most particles have not had a single pericenter passage. There are some fractions of particles with greater than a few passages (between 2 and 4); these particles mostly correlate with regions of high pericenter passage distance. The right panel on Figure \ref{fig:1d-distributions} shows the distribution of radial velocities, $V_r$ normalized by $v_{\rm cmax}$ of the host halo. The particles inside the ellipse form a nearly Gaussian distribution that is expected in a thermalized distribution of velocities.

\subsection{Separation of Substructures}
While the ellipse, i.e., the ellipse in UMAP space, corresponds to $68.5\%$ of the primary host halo particle, it is interesting to see how well the UMAP representation separates out subhalos and other substructures from within the host. The larger question is whether the substructure occupies distinct regions in the 6D phase space and whether they are separable from the host particles using such Machine Learning techniques. A priori, we note that the ellipse is largely smooth, even though some subhalos do exist within; these are typical of mass $\le 2.3 \times 10^{9} \Smass$. Using the subhalo histories extracted from ROCKSTAR, we find that the subhalo centers that are mapped to within the ellipse, using the transformation provided by UMAP, $\Phi$ (described in section \ref{sec:UMAP}) have mostly completed several orbits within the halo. All have had at least one pericenter passage, and most have had two or above. Most of the large infalling objects are clearly placed outside the largest ellipse.

Additionally, at $z=0$, we also post-process the particles within the ellipse with DBSCAN. We extract all the particles within the ellipse and run DBSCAN on their location in 6D space. This is to identify structures that appear dynamically singular with each other in UMAP space. The results are shown in Fig. \ref{fig:Ellipse_DBSCAN}. The points on the left are all the cluster points that DBSCAN finds in the radial phase space. Note that these are distinct from the objects identified \textit{outside} the ellipse, even though they occupy similar regions in the radial phase space (Figure \ref{fig:phasespace_log}). The right panel shows the ``noise'' points identified by DBSCAN. This corresponds to $95\%$ of the points within the ellipse and depicts the smooth component of the halo. The streams/caustic-like structures are clearly visible in this space. However, it is non-trivial to identify if these are truly dark matter caustics like those described in \cite{FG84}, or distinct stream-like structures created by disrupted substructure. Overall, we find that while most of the ellipse is smooth, the particles that are assigned to the clusters typically have had about 2 pericenter passages. This number falls steeply beyond 2 with very few clustered particles that have had more than 4 pericenter passages, showing that most substructures do not persist beyond 4 pericenter passages as coherent structures.

Typical subhalos falling into the host are shown in Fig. \ref{fig:subhalos_umap}. To isolate the subhalo particles, we have identified a subhalo at $z=0$ from the ROCKSTAR catalog, traced it back to a redshift before infall, and selected particles within $1.5 r_{\rm vir}$ of the subhalo at the time. The leftmost panel shows a subhalo in the UMAP space. The middle panel shows the subhalo in $x-y$ space, and the right panel in radial phase space. We run DBSCAN in UMAP space to separate the clumped features and color them distinctly. The top row shows a subhalo of mass $6.4\times 10^9 \Smass$. We note that the primary subhalo core and a part of its leading and trailing arm, which is in close proximity with it in the radial phase space, are placed outside the ellipse (dark blue points), whereas green particles that entered with the subhalo have been dislodged from the subhalo and do not seem to be associated with the original subhalo any more. 

We have seen that UMAP separates a single subhalo into multiple components based on the different dynamics exhibited by the particles of the subhalo. We also try to identify clusters in UMAP space using DBSCAN and study the cluster mass function. The results are described in Appendix \ref{app:mass_func}. While the cluster mass function appears to have the correct slope and features of the halo mass function in the Milky Way field, we find that some of the structures picked out in the UMAP space do not necessarily correspond to halo-like structures in real space. We therefore only use this as a measure of mass in clustered structure in the UMAP space, we defer a detailed analysis of identifying halos in UMAP space to a later work. 

\begin{figure*}
    \centering
    \includegraphics[width=0.9\textwidth]{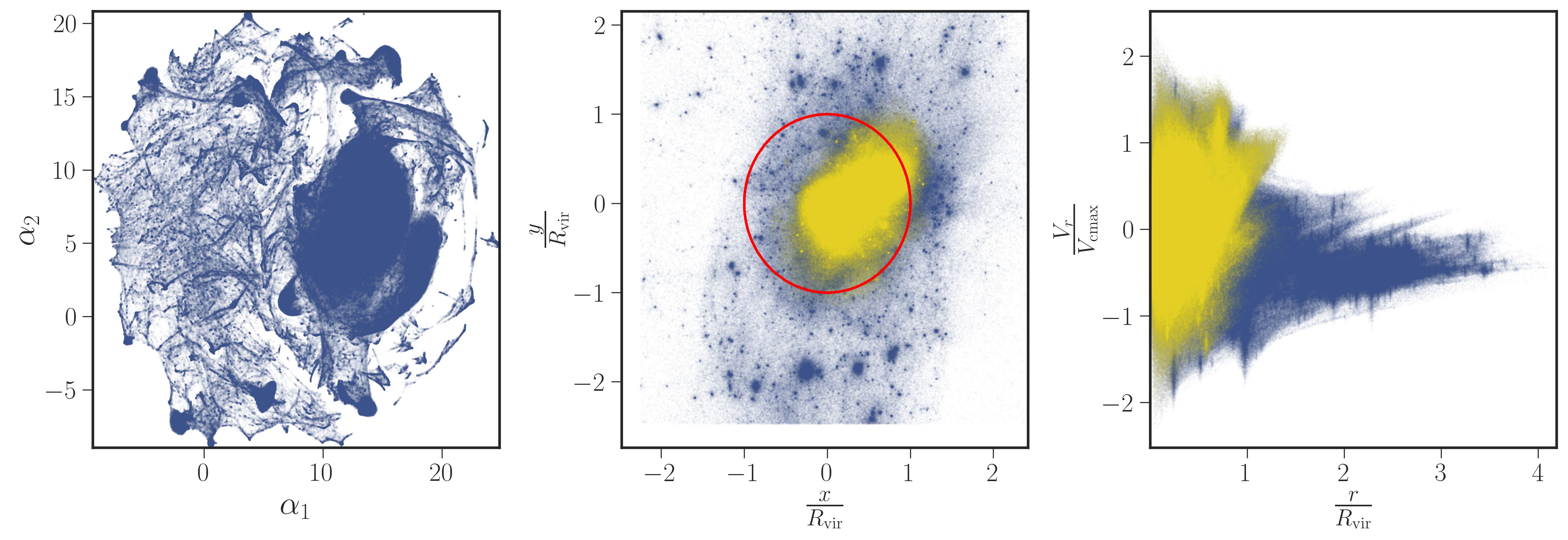}
    \caption{\textbf{Left : } UMAP representation of a $1\Mpch$ box centerd around the MW at $z=1.0$. The yellow points correspond to the most massive cluster identified by DBSCAN.
    \textbf{Center : }shows the cluster from UMAP space (yellow) plotted in the $x-y$ plane against the rest of the particles in the box (deep blue). The red solid circle is at the virial radius of the Milky Way at $z=1.0$. \textbf{Right : }shows the radial phase space ($V_r-r$) plot of the same cluster (yellow) against the phase space distribution of the entire box (deep blue). The halo is undergoing a merger at this redshift, making the UMAP representation of the largest cluster irregular.}
    \label{fig:Ellipse_z1}
\end{figure*}

\subsection{Redshift Evolution and dynamical mergers in UMAP space}

Our fiducial redshift for analysis is $z=0$, and the half-mass timescale of the fiducial Milky Way-like halo is at $z=1.38$. The half-mass timescale is an estimate of the formation time of the halo. Our halo has an age that is slightly older than the mean age of Milky Way like halos \citep{2002ApJ...568...52W} and is a relatively old system (as halos at these masses typically are compared to massive groups and clusters). We expect such a system to be largely relaxed at the present time. Earlier in cosmic time, in the halo's history, however, it is known that the halo went through several major mergers and a period of fast accretion where the structure within the halo undergoes rapid changes \citep{Wechsler:2005gb, Wang:2020hpl}.  

We run the UMAP algorithm on a $1\Mpch$ region around the Milky Way center at each snapshot with a fixed $\nn$ value of 30. The earliest redshift we have saved in the simulation is $z=19$. The field is usually dominated by a single large cluster as in the $z=0$ case after approximately $z=2.2$. During the early stages of the evolution, the largest single cluster in the UMAP field is not well approximated by a single ellipse. We note that this is exceptionally visible during the phase of major mergers. Figure \ref{fig:Ellipse_z1} shows the UMAP representation and the largest cluster identified by DBSCAN in UMAP space at $z=1$. We pick this specific snapshot as the halo is going through a merger here. The largest cluster in the UMAP space does not appear to be a well-formed elliptical region in this space, depicting signatures of an ongoing merger. The middle and right-hand panels of the same figure show the particles associated with the largest cluster identified using DBSCAN in the UMAP space overlaid on the configuration space and radial phase space of the Milky Way. The dynamically distinct region still largely corresponds to the orbiting particles that are in the process of mixing.

\section{Discussion $\&$ Conclusion}
\label{sec:conclusions}

In this paper, we have explored the structures within a Milky Way mass dark matter halo using a combination of an unsupervised machine learning algorithm, Uniform Manifold Approximation and Projection for Dimension Reduction (UMAP), and a clustering algorithm, Density-Based Spatial Clustering of Applications with Noise (DBSCAN). UMAP is an efficient way of separating correlated structures that form groups in higher dimensional space, and representing them in a lower dimensional space while preserving topological information. DBSCAN serves the purpose of clustering in the output space produced by UMAP. 

We find that the dark matter particles in a $1\Mpch$ region around a Milky Way-like halo ($R_{\rm vir}=0.2\Mpch$) clearly separate out into distinct structures in the UMAP space, with the bulk of the particles within the halo radius forming the largest and most dominant cluster in this space. In the specific choice of UMAP parameters, the largest cluster appears like a nearly uniform-density ellipse. This ellipse, however, contains only $68.5\%$ of the dark matter particles that are within the host virial radius, implying that even within the halo radius, the particles are not all dynamically uniform. Comparing the particle UMAP locations to their location in real space and in the radial phase space, we note that the particles in the ellipse in UMAP space are confined to $\sim 80\%$ of the virial radius and do not include the infall stream or even the particles near the splashback (first apocenter of the most recently accreted particles) region of the halo. 

To interpret the UMAP representation, we correlate the UMAP positions with the history/orbital properties of the particles. We find that the largest cluster in UMAP space mostly contains particles that have had multiple pericenter passages (Fig. \ref{fig:pericenter}). In fact, a sharp peak appears in the distribution of $n_{\rm orbit}$ of the particles within the ellipse at $n_{\rm orbit}=4$. These are particles that are, therefore, expected to have been phase-mixed and are dynamically separated from the more recently accreted halo, which is placed outside the boundary of the cluster. 

UMAP representation clearly shows that the halo is not a uniform, virialized structure and retains information about evolutionary history, most prominently, the recently accreted particles, which form a distinct set of particles from the older halo that is the largest ellipse. Even particles that are currently at splashback or were at splashback roughly a dynamical time before today, do not occupy spaces in the UMAP representation that correspond to the ``relaxed halo''. We also find that particles with higher angular momentum (large pericenters) are almost all placed outside the largest cluster. Therefore, overall, the number of orbits and the angular momentum in those orbits play an important role in the current dynamic state of a particle within a halo. UMAP acts as a powerful tool to separate out the dynamic structures within the halo \textit{without using any historical information or any a priori assumptions about the state of the halo}. Our results are also consistent with \cite{Lucie-Smith:2022mar} and \cite{Shin:2022iza}, where they correlate current profiles and the accretion history of dark matter halos to find that the outer profiles of halos are strongly dependent on the late accretion history. UMAP representations provide a simple, intuitive way to separate these regions with phase space information today.

We use UMAP as a pre-processing step to find clusters using DBSCAN to test its usability as a halo finder. With the DBSCAN algorithm optimized to find structures within the larger ellipse, we find that it neatly separates out the smooth and clustered components. However, the vast diversity of features outside the ellipse makes it difficult to identify single halos as one cluster, we will defer a detailed investigation to optimize halo finding in this space to future work.

We also project the halo field at different redshifts of the simulation and find that an algorithm like UMAP is ideal for studying the evolution of the dynamical state of the halo and its surroundings. We find that with time, the field is dominated by a single large cluster that evolves towards a smooth ellipse, with several phases of persisting irregular shapes during major mergers. The approach towards the smoother ellipse is an evolution towards relaxation of the central region of the halo. 

Our work shows halos are evolving, non-uniform structures in phase space. Even for a fairly early forming object, with no surviving subhalos that has a mass more than $0.01 M_{\rm host}$ for significant gravitational perturbations, the internal structures are not completely phase mixed. Galaxy formation is intimately tied to the dynamical assembly of the halo. It will be important to understand how the properties of galaxies that live in (or are accreted in) the different regions in the UMAP space vary in their stellar mass, star-formation, and morphology properties \citep{More:2015ufa}. Further, it will be useful to use this technique in larger cosmological volumes and study the inter-relation between halos and large-scale structures and embed galaxy formation models based on UMAP weights. Recently, \cite{Salazar:2024ifw} have also shown that separately addressing the orbiting and infalling particles within halos while modeling the halo term is a promising and robust approach to fit the full non-linear matter correlation function. It will be interesting to separate halo regions based on UMAP representations to understand their impact on modeling such terms both in configuration and redshift space.

UMAP is a simple, easy-to-use projection technique that allows us to visualize different dynamical structures in higher dimensional space. Since UMAP is a machine-learning algorithm, along with producing a low-dimensional representation, UMAP also produces a mapping function between the input and output spaces. The viability of using $\Phi$ for a different box/simulation without re-training will be an important extension to the current work that will likely produce further insight into halo formation and evolution.

\section*{Acknowledgements}
We thank Arka Banerjee, Chihway Chang,  Ranjan Laha, and Judit Prat for their useful discussions. We received support from the PARAM Brahma supercomputing facility at IISER Pune, which is part of the National Super Computing Mission under the Government of India.  This project was supported under SERB grant SRG/2023/001563. The authors thank the organizers of the Cosmo21 meeting for useful discussions related to the use of Artificial Intelligence and Machine learning in Cosmology.

\bibliographystyle{yahapj}
\bibliography{references}

\begin{thebibliography}{}
\providecommand\natexlab[1]{#1}
\providecommand\JournalTitle[1]{#1}

\bibitem[{Adhikari {et~al.}(2014)Adhikari, Dalal, \& Chamberlain}]{Adhikari:2014lna}
Adhikari, S., Dalal, N., \& Chamberlain, R.~T. 2014, \href{http://dx.doi.org/10.1088/1475-7516/2014/11/019}{\JournalTitle{JCAP}, 11, 019}

\bibitem[{Angulo \& Hahn(2021)}]{Angulo:2021kes}
Angulo, R.~E. \& Hahn, O. \href{http://dx.doi.org/10.1007/s41115-021-00013-z}{2021}, \href{http://arxiv.org/abs/2112.05165}{{\sffamily arXiv:2112.05165 [astro-ph.CO]}}

\bibitem[{Aung {et~al.}(2021)Aung, Nagai, Rozo, \& Garcia}]{Aung:2020czp}
Aung, H., Nagai, D., Rozo, E., \& Garcia, R. 2021, \href{http://dx.doi.org/10.1093/mnras/staa3994}{\JournalTitle{Mon. Not. Roy. Astron. Soc.}, 502, 1041}

\bibitem[{Bagla(2005)}]{Bagla:2004au}
Bagla, J.~S. 2005, \JournalTitle{Curr. Sci.}, 88, 1088

\bibitem[{Bahcall {et~al.}(1999)Bahcall, Ostriker, Perlmutter, \& Steinhardt}]{Bahcall_1999}
Bahcall, N.~A., Ostriker, J.~P., Perlmutter, S., \& Steinhardt, P.~J. 1999, \href{http://dx.doi.org/10.1126/science.284.5419.1481}{\JournalTitle{Science}, 284, 1481–1488}

\bibitem[{{Bardeen} {et~al.}(1986){Bardeen}, {Bond}, {Kaiser}, \& {Szalay}}]{1986ApJ...304...15B}
{Bardeen}, J.~M., {Bond}, J.~R., {Kaiser}, N., \& {Szalay}, A.~S. 1986, \href{http://dx.doi.org/10.1086/164143}{\JournalTitle{\apj}, 304, 15}

\bibitem[{Behroozi {et~al.}(2012{\natexlab{a}})Behroozi, Wechsler, \& Wu}]{Behroozi_2012A}
Behroozi, P.~S., Wechsler, R.~H., \& Wu, H.-Y. 2012{\natexlab{a}}, \href{http://dx.doi.org/10.1088/0004-637x/762/2/109}{\JournalTitle{The Astrophysical Journal}, 762, 109}

\bibitem[{Behroozi {et~al.}(2012{\natexlab{b}})Behroozi, Wechsler, Wu, Busha, Klypin, \& Primack}]{Behroozi_2012B}
Behroozi, P.~S., Wechsler, R.~H., Wu, H.-Y., {et~al.} 2012{\natexlab{b}}, \href{http://dx.doi.org/10.1088/0004-637x/763/1/18}{\JournalTitle{The Astrophysical Journal}, 763, 18}

\bibitem[{Belkin \& Niyogi(2001)}]{10.5555/2980539.2980616}
Belkin, M. \& Niyogi, P. 2001, in Proceedings of the 14th International Conference on Neural Information Processing Systems: Natural and Synthetic, NIPS'01 (Cambridge, MA, USA: MIT Press), 585–591

\bibitem[{Binney \& Tremaine(2008)}]{binney_tremaine_2008}
Binney, J. \& Tremaine, S. 2008, Galactic Dynamics (Princeton University Press)

\bibitem[{{Blumenthal} {et~al.}(1984){Blumenthal}, {Faber}, {Primack}, \& {Rees}}]{1984Natur.311..517B}
{Blumenthal}, G.~R., {Faber}, S.~M., {Primack}, J.~R., \& {Rees}, M.~J. 1984, \href{http://dx.doi.org/10.1038/311517a0}{\JournalTitle{\nat}, 311, 517}

\bibitem[{Cooray \& Sheth(2002)}]{COORAY_2002}
Cooray, A. \& Sheth, R. 2002, \href{http://dx.doi.org/10.1016/s0370-1573(02)00276-4}{\JournalTitle{Physics Reports}, 372, 1–129}

\bibitem[{Cuesta {et~al.}(2008)Cuesta, Prada, Klypin, \& Moles}]{Cuesta:2007it}
Cuesta, A.~J., Prada, F., Klypin, A., \& Moles, M. 2008, \href{http://dx.doi.org/10.1111/j.1365-2966.2008.13590.x}{\JournalTitle{Mon. Not. Roy. Astron. Soc.}, 389, 385}

\bibitem[{{Davis} {et~al.}(1985){Davis}, {Efstathiou}, {Frenk}, \& {White}}]{1985ApJ...292..371D}
{Davis}, M., {Efstathiou}, G., {Frenk}, C.~S., \& {White}, S.~D.~M. 1985, \href{http://dx.doi.org/10.1086/163168}{\JournalTitle{\apj}, 292, 371}

\bibitem[{{Diemer}(2017)}]{2017ApJS..231....5D}
{Diemer}, B. 2017, \href{http://dx.doi.org/10.3847/1538-4365/aa799c}{\JournalTitle{\apjs}, 231, 5}

\bibitem[{{Diemer}(2020)}]{2020ApJS..251...17D}
{Diemer}, B. 2020, \href{http://dx.doi.org/10.3847/1538-4365/abbf51}{\JournalTitle{\apjs}, 251, 17}

\bibitem[{Diemer \& Kravtsov(2014)}]{Diemer:2014xya}
Diemer, B. \& Kravtsov, A.~V. 2014, \href{http://dx.doi.org/10.1088/0004-637X/789/1/1}{\JournalTitle{Astrophys. J.}, 789, 1}

\bibitem[{{Efstathiou} {et~al.}(1985){Efstathiou}, {Davis}, {White}, \& {Frenk}}]{1985ApJS...57..241E}
{Efstathiou}, G., {Davis}, M., {White}, S.~D.~M., \& {Frenk}, C.~S. 1985, \href{http://dx.doi.org/10.1086/191003}{\JournalTitle{\apjs}, 57, 241}

\bibitem[{Errani \& Navarro(2021)}]{Errani_2021}
Errani, R. \& Navarro, J.~F. 2021, \href{http://dx.doi.org/10.1093/mnras/stab1215}{\JournalTitle{Monthly Notices of the Royal Astronomical Society}, 505, 18–32}

\bibitem[{Fillmore \& Goldreich(1984)}]{FG84}
Fillmore, J.~A. \& Goldreich, P. 1984, \href{http://dx.doi.org/10.1086/162070}{\JournalTitle{Astrophys. J.}, 281, 1}

\bibitem[{Garcia {et~al.}(2023)Garcia, Salazar, Rozo, Adhikari, Aung, Diemer, Nagai, \& Wolfe}]{Garcia:2022zsz}
Garcia, R., Salazar, E., Rozo, E., {et~al.} 2023, \href{http://dx.doi.org/10.1093/mnras/stad660}{\JournalTitle{Mon. Not. Roy. Astron. Soc.}, 521, 2464}

\bibitem[{Green \& van den Bosch(2019)}]{Green_2019}
Green, S.~B. \& van den Bosch, F.~C. 2019, \href{http://dx.doi.org/10.1093/mnras/stz2767}{\JournalTitle{Monthly Notices of the Royal Astronomical Society}, 490, 2091–2101}

\bibitem[{{Gunn} \& {Gott}(1972)}]{1972ApJ...176....1G}
{Gunn}, J.~E. \& {Gott}, J.~Richard, I. 1972, \href{http://dx.doi.org/10.1086/151605}{\JournalTitle{\apj}, 176, 1}

\bibitem[{{Hahn} \& {Abel}(2011)}]{2011MNRAS.415.2101H}
{Hahn}, O. \& {Abel}, T. 2011, \href{http://dx.doi.org/10.1111/j.1365-2966.2011.18820.x}{\JournalTitle{\mnras}, 415, 2101}

\bibitem[{Hayashi {et~al.}(2003)Hayashi, Navarro, Taylor, Stadel, \& Quinn}]{Hayashi_2003}
Hayashi, E., Navarro, J.~F., Taylor, J.~E., Stadel, J., \& Quinn, T. 2003, \href{http://dx.doi.org/10.1086/345788}{\JournalTitle{The Astrophysical Journal}, 584, 541–558}

\bibitem[{{Klypin} {et~al.}(1999){Klypin}, {Gottl{\"o}ber}, {Kravtsov}, \& {Khokhlov}}]{1999ApJ...516..530K}
{Klypin}, A., {Gottl{\"o}ber}, S., {Kravtsov}, A.~V., \& {Khokhlov}, A.~M. 1999, \href{http://dx.doi.org/10.1086/307122}{\JournalTitle{\apj}, 516, 530}

\bibitem[{{Klypin} \& {Shandarin}(1983)}]{1983MNRAS.204..891K}
{Klypin}, A.~A. \& {Shandarin}, S.~F. 1983, \href{http://dx.doi.org/10.1093/mnras/204.3.891}{\JournalTitle{\mnras}, 204, 891}

\bibitem[{{Knollmann} \& {Knebe}(2009)}]{2009ApJS..182..608K}
{Knollmann}, S.~R. \& {Knebe}, A. 2009, \href{http://dx.doi.org/10.1088/0067-0049/182/2/608}{\JournalTitle{\apjs}, 182, 608}

\bibitem[{Lithwick \& Dalal(2011)}]{LD10}
Lithwick, Y. \& Dalal, N. 2011, \href{http://dx.doi.org/10.1088/0004-637X/734/2/100}{\JournalTitle{Astrophys. J.}, 734, 100}

\bibitem[{Lucie-Smith {et~al.}(2022)Lucie-Smith, Adhikari, \& Wechsler}]{Lucie-Smith:2022mar}
Lucie-Smith, L., Adhikari, S., \& Wechsler, R.~H. 2022, \href{http://dx.doi.org/10.1093/mnras/stac1833}{\JournalTitle{Mon. Not. Roy. Astron. Soc.}, 515, 2164}

\bibitem[{Mansfield {et~al.}(2017)Mansfield, Kravtsov, \& Diemer}]{Mansfield:2016bxx}
Mansfield, P., Kravtsov, A.~V., \& Diemer, B. 2017, \href{http://dx.doi.org/10.5281/zenodo.569034}{\JournalTitle{Astrophys. J.}, 841, 34}

\bibitem[{Mao {et~al.}(2015)Mao, Williamson, \& Wechsler}]{Mao_2015}
Mao, Y.-Y., Williamson, M., \& Wechsler, R.~H. 2015, \href{http://dx.doi.org/10.1088/0004-637x/810/1/21}{\JournalTitle{The Astrophysical Journal}, 810, 21}

\bibitem[{McInnes {et~al.}(2020)McInnes, Healy, \& Melville}]{mcinnes2020umapuniformmanifoldapproximation}
McInnes, L., Healy, J., \& Melville, J. 2020, UMAP: Uniform Manifold Approximation and Projection for Dimension Reduction, \href{http://arxiv.org/abs/1802.03426}{{\sffamily arXiv:1802.03426 [stat.ML]}}

\bibitem[{More {et~al.}(2015)More, Diemer, \& Kravtsov}]{More:2015ufa}
More, S., Diemer, B., \& Kravtsov, A. 2015, \href{http://dx.doi.org/10.1088/0004-637X/810/1/36}{\JournalTitle{Astrophys. J.}, 810, 36}

\bibitem[{Navarro {et~al.}(1997)Navarro, Frenk, \& White}]{Navarro:1996gj}
Navarro, J.~F., Frenk, C.~S., \& White, S. D.~M. 1997, \href{http://dx.doi.org/10.1086/304888}{\JournalTitle{Astrophys. J.}, 490, 493}

\bibitem[{Peñarrubia {et~al.}(2010)Peñarrubia, Benson, Walker, Gilmore, McConnachie, \& Mayer}]{Pe_arrubia_2010}
Peñarrubia, J., Benson, A.~J., Walker, M.~G., {et~al.} 2010, \href{http://dx.doi.org/10.1111/j.1365-2966.2010.16762.x}{\JournalTitle{Monthly Notices of the Royal Astronomical Society}, no}

\bibitem[{Salazar {et~al.}(2024)Salazar, Rozo, Garc\'\i{}a, Kokron, Adhikari, Diemer, \& Osinga}]{Salazar:2024ifw}
Salazar, E.~M., Rozo, E., Garc\'\i{}a, R., {et~al.} 2024, \href{http://arxiv.org/abs/2406.04054}{{\sffamily arXiv:2406.04054 [astro-ph.CO]}}

\bibitem[{Shi(2016)}]{Shi:2016lwp}
Shi, X. 2016, \href{http://dx.doi.org/10.1093/mnras/stw925}{\JournalTitle{Mon. Not. Roy. Astron. Soc.}, 459, 3711}

\bibitem[{Shin \& Diemer(2023)}]{Shin:2022iza}
Shin, T.-h. \& Diemer, B. 2023, \href{http://dx.doi.org/10.1093/mnras/stad860}{\JournalTitle{Mon. Not. Roy. Astron. Soc.}, 521, 5570}

\bibitem[{{Springel} {et~al.}(2001){Springel}, {White}, {Tormen}, \& {Kauffmann}}]{2001MNRAS.328..726S}
{Springel}, V., {White}, S. D.~M., {Tormen}, G., \& {Kauffmann}, G. 2001, \href{http://dx.doi.org/10.1046/j.1365-8711.2001.04912.x}{\JournalTitle{\mnras}, 328, 726}

\bibitem[{{Tomita}(1969)}]{1969PThPh..42....9T}
{Tomita}, K. 1969, \href{http://dx.doi.org/10.1143/PTP.42.9}{\JournalTitle{Progress of Theoretical Physics}, 42, 9}

\bibitem[{Wang {et~al.}(2020)Wang, Mao, Zentner, Lange, van~den Bosch, \& Wechsler}]{Wang:2020hpl}
Wang, K., Mao, Y.-Y., Zentner, A.~R., {et~al.} 2020, \href{http://dx.doi.org/10.1093/mnras/staa2733}{\JournalTitle{Mon. Not. Roy. Astron. Soc.}, 498, 4450}

\bibitem[{{Wechsler} {et~al.}(2002){Wechsler}, {Bullock}, {Primack}, {Kravtsov}, \& {Dekel}}]{2002ApJ...568...52W}
{Wechsler}, R.~H., {Bullock}, J.~S., {Primack}, J.~R., {Kravtsov}, A.~V., \& {Dekel}, A. 2002, \href{http://dx.doi.org/10.1086/338765}{\JournalTitle{\apj}, 568, 52}

\bibitem[{Wechsler \& Tinker(2018)}]{Wechsler:2018pic}
Wechsler, R.~H. \& Tinker, J.~L. 2018, \href{http://dx.doi.org/10.1146/annurev-astro-081817-051756}{\JournalTitle{Ann. Rev. Astron. Astrophys.}, 56, 435}

\bibitem[{Wechsler {et~al.}(2006)Wechsler, Zentner, Bullock, \& Kravtsov}]{Wechsler:2005gb}
Wechsler, R.~H., Zentner, A.~R., Bullock, J.~S., \& Kravtsov, A.~V. 2006, \href{http://dx.doi.org/10.1086/507120}{\JournalTitle{Astrophys. J.}, 652, 71}

\bibitem[{{White} \& {Rees}(1978)}]{1978MNRAS.183..341W}
{White}, S.~D.~M. \& {Rees}, M.~J. 1978, \href{http://dx.doi.org/10.1093/mnras/183.3.341}{\JournalTitle{\mnras}, 183, 341}

\bibitem[{Zhou \& Han(2023)}]{Zhou:2023uhb}
Zhou, Y. \& Han, J. 2023, \href{http://dx.doi.org/10.1093/mnras/stad2375}{\JournalTitle{Mon. Not. Roy. Astron. Soc.}, 525, 2489}

\end{thebibliography}

\appendix
\counterwithin{figure}{section}

\section{UMAP and its Basic Hyper-Parameters}
\begin{figure*}
    \centering
    \includegraphics[width=0.9\textwidth]{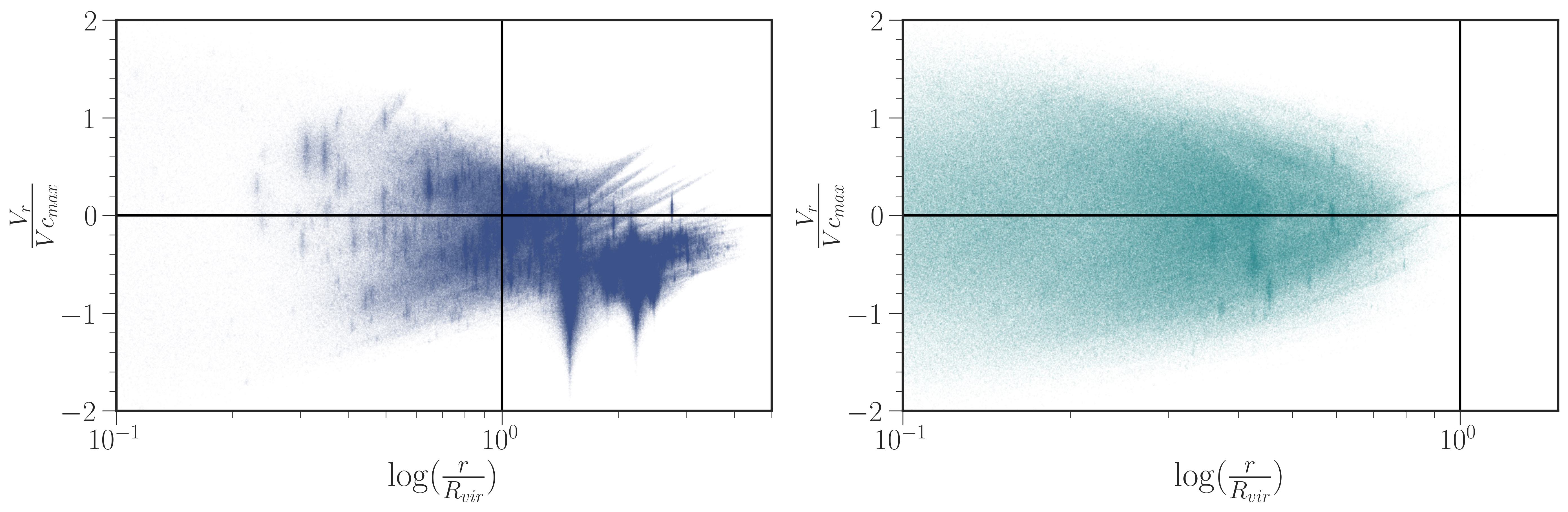}
    \includegraphics[width=0.9\textwidth]{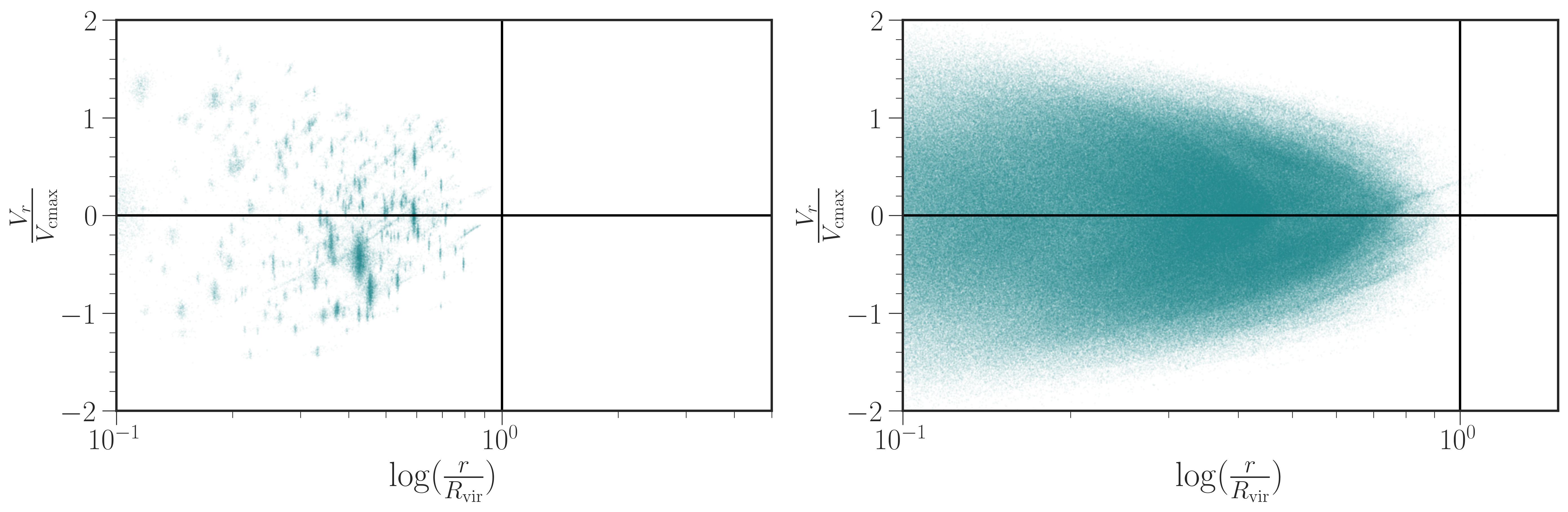}
    \caption{\textbf{Top row : }Radial phase space scatter plot in log space.
    \textit{Left :} Radial phase space location of particles that are placed outside the ellipse in the UMAP space. \textit{Right :} Locations of the particles which are placed inside the ellipse (see Figure \ref{fig:Ellipse} for reference). \textbf{Bottom row : }DBSCAN clustering run on the 6D information of the particles inside the ellipse from Figure \ref{fig:Ellipse} to isolate substructure.\textit{Left :} shows the particles assigned to clusters by DBSCAN. \textit{Right : } shows the particles classified as noise points by DBSCAN. $\sim 94.3\%$ of the particles are classified as noise, implying that most of the particles in the largest ellipse form the smooth component of the halo.}\label{fig:appendix_ellipse_logspace}
\end{figure*}

\subsection{UMAP - The Algorithm} \label{app:umap}
This part is a paraphrased version of the mathematics presented in \cite{mcinnes2020umapuniformmanifoldapproximation}. Uniform Manifold Approximation and Projection for Dimension Reduction (UMAP) is a manifold learning technique based on Riemannian geometry and algebraic topology. From a computational point of view, UMAP simply constructs and manipulates weighted graphs. This puts UMAP in the category of k-neighbor-based graph learning algorithms. UMAP assumes these three conditions to be axiomatically true
\begin{enumerate}
    \item There exists a manifold on which the data would be uniformly distributed 
    \item The underlying manifold of interest is locally connected
    \item Preserving the topological structure of the manifold is the primary goal
\end{enumerate}
UMAP can be broken down into four broad steps.
\begin{enumerate}
    \item Generate a weighted graph. This will be the source graph. Referred to as $\mathcal{G}$ in text.
    \item Initialize a low dimensional (target dimension, provided by the user) graph using spectral embedding. A random initialization works in theory, but spectral embedding converges better and faster.
    \item Generate a weighted graph for the low-dimensional embedding. Referred to as $\mathcal{H}$ in text.
    \item Use a force-directed graph layout algorithm to optimize the low-dimensional weighted graph to resemble the source graph as closely as practically possible. This, in a sense, preserves the topology.
\end{enumerate}

\subsubsection{Graph Construction}
Let $X = \{x_1, x_2, ... x_N\}$ be the input dataset and $d$ be the metric in this space. Given an input parameter $k$, for each $x_i$, UMAP computes the set $\{x_{i_1}, x_{i_2}, ...,x_{i_k}\}$ of the $k$ nearest neighbors of $x_i$ under the metric $d$. Once this is obtained, UMAP calculates $\rho_i$ and $\sigma_i$ for each $x_i$ according to the following equations
\begin{equation}
    \rho_i = {\rm min}\{d(x_i,x_{i_j}) \thickspace | \thickspace 1\leq j \leq k, \thickspace d(x_i, x_{i_j}) > 0\}
\end{equation}
\begin{equation}
     \sum_{j=1}^{k} \exp\left(\frac{- \thickspace {\rm max}[0,\thickspace d(x_i,x_{i_j}) \thinspace - \thinspace \rho_i]}{\sigma_i}\right)\thickspace = \thickspace \log_2(k) 
\end{equation}
UMAP then defines the weighted graph using $\mathcal{G} = (V,E,w)$ where $V$ is simply $X$, 
\begin{equation}
    E \thickspace = \thickspace \{(x_i, x_{i_j}) \thickspace | \thickspace 1\leq j \leq k, \thickspace 1\leq i \leq N\} {\rm, and} 
\end{equation}
\begin{equation}
    w((x_i, x_{i_j})) = \exp\left(\frac{-\thickspace {\rm max} [0,\thickspace d(x_i,x_{i_j}) \thinspace - \thinspace \rho_i]}{\sigma_i}\right)
\end{equation}
This would imply that the edge between two fixed points $x_i$ and $x_j$ would have two different weights for the two directions depending on the distribution of points in the neighborhood of each of the points. To combine them to form a unified topological representation, let's look at $A$, the weighted adjacent matrix of $\mathcal{G}$, and consider the symmetric matrix 
\begin{equation}
    B = A + A^T - A \circ A^T
\end{equation}
where $\circ$ is the Hadamard (or pointwise) product. Then, a graph $\mathcal{G}$ is an undirected weighted graph whose adjacency matrix is given by $B$. 
\begin{figure*}
    \centering
    \includegraphics[width=0.9\linewidth]{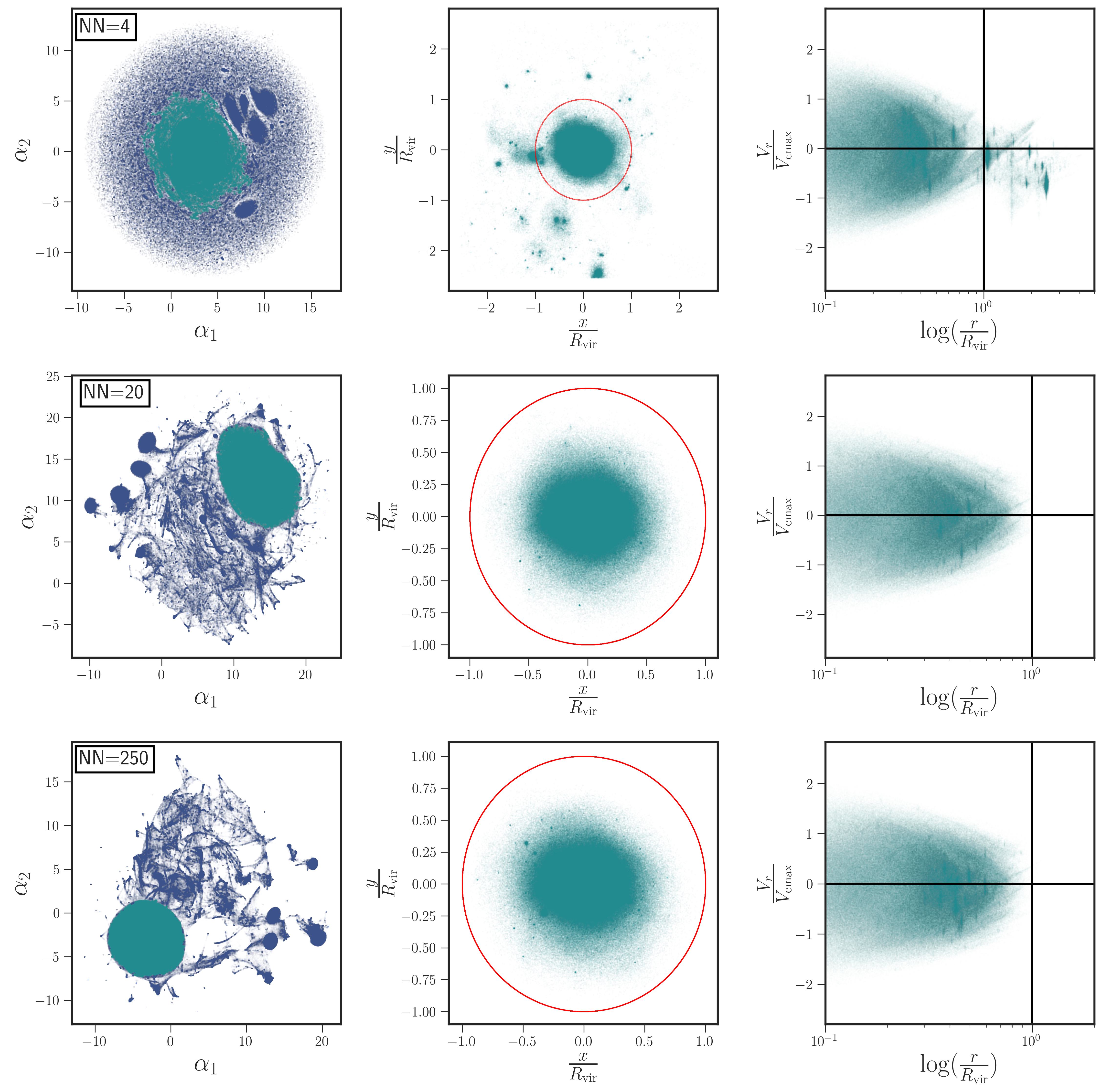}
    \caption{\textbf{Left column : }UMAP reduced 2D scatter plot of the $1\Mpch$ box centerd around the MW at $z=0$ for varying values of $\nn$. The $\nn$ values are indicated in the respective plots. The rest of the parameters are kept constant. The colored region corresponds to the largest cluster identified by DBSCAN. DBSCAN classified $4\%, 10\% $ and $13\%$ of the points as noise for $nn = 4$, $20$ and $250$ respectively. \textbf{Middle column : }shows the largest identified cluster in real space. The red circle is the virial boundary of the MW. \textbf{Right column : }shows the $V_r-r$ space distribution of the largest identified cluster.}
    \label{fig:appendix_umap_nn}
\end{figure*}

\subsubsection{Graph Layout}
A force-directed approach makes use of attractive forces applied along edges, and repulsive forces applied on vertices. The attractive force between two vertices $i$ and $j$ at low-dimensional coordinates $y_i$ and $y_j$, respectively, is determined by 
\begin{equation}
    \frac{-2ab \thinspace ||y_i \thinspace - \thinspace y_j||^{2(b - 1)}_2}{1 \thinspace + \thinspace ||y_i \thinspace - \thinspace y_j||^2_2}\thickspace w((x_i,x_j)) \thickspace (y_i \thinspace - \thinspace y_j)
\end{equation}
where $a$ and $b$ are hyper-parameters.
Repulsive forces are computed via sampling due to computational constraints. Thus, whenever an attractive force is applied between two vertices $i$ and $j$, one of them is repulsed from some other vertex $k$, chosen via sampling. The repulsive force is given by 
\begin{equation}
    \frac{2b}{(\epsilon \thinspace + \thinspace ||y_i \thinspace - \thinspace y_j||_2^2)\thickspace(1 \thinspace + \thinspace a ||y_i \thinspace - \thinspace y_j||^{2b}_2)}\thickspace (1 \thinspace - \thinspace w((x_i,x_j))) \thickspace (y_i \thinspace - \thinspace y_j)
\end{equation}
$\epsilon$ is a small number (0.001) to avoid division by zero.\\
These forces are derived gradients optimizing the edge-wise cross-entropy between the weighted source graph $\mathcal{G}$ and an equivalent weighted graph $\mathcal{H}$ generated using $\{y_i\}_{i=1\dots N}$. Therefore, $\{y_i\}$ is transformed such that the cross entropy loss between $\mathcal{H}$ and $\mathcal{G}$ is minimum, i.e., topology is conserved in the low-dimension representation.

\begin{figure}
    \centering
    \includegraphics[width=\linewidth]{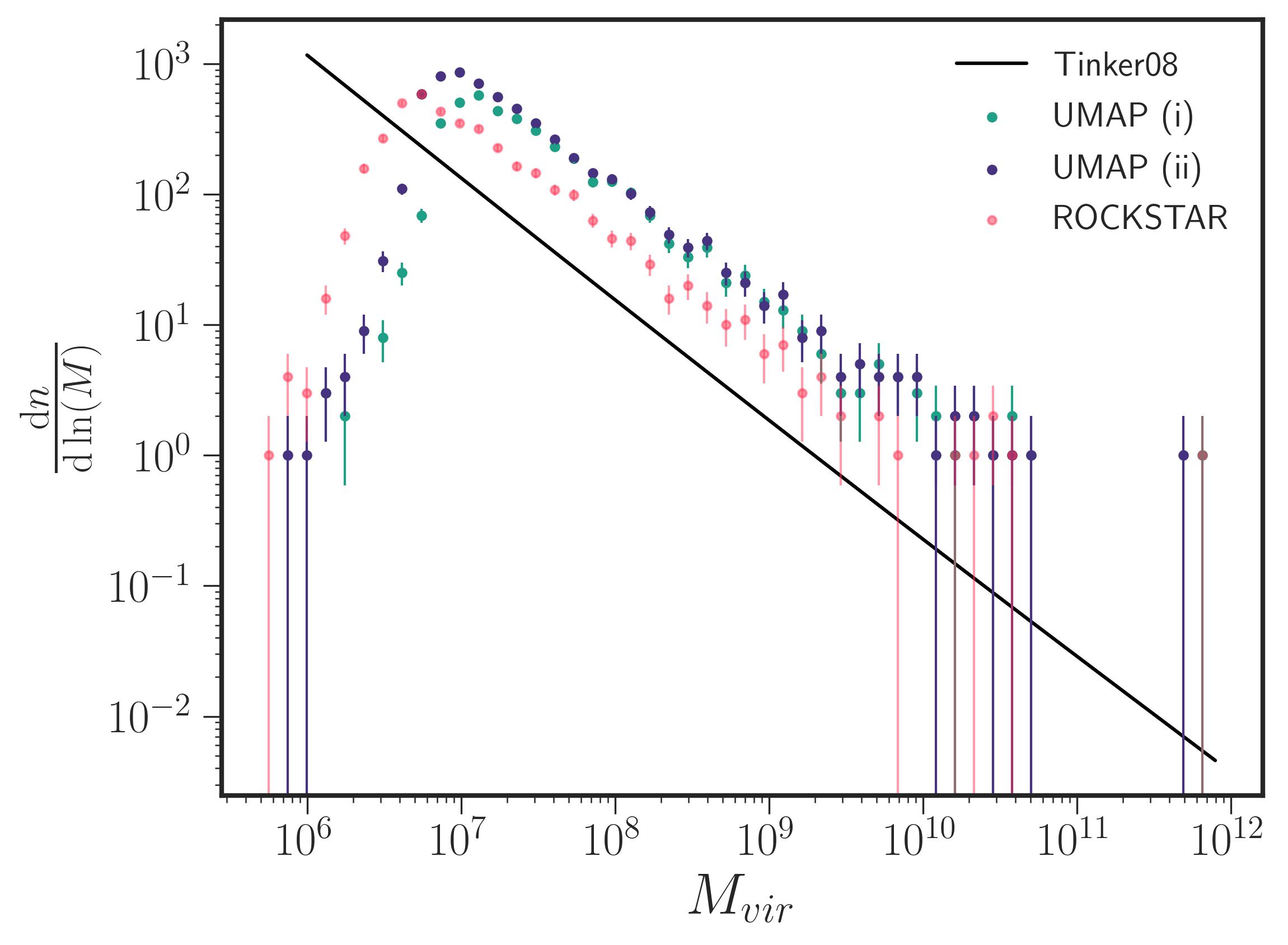}
    \caption{The solid line represents the Tinker halo mass function at $z=0$ for halos of mass in the range $10^{6}-10^{12} \Smass$. The deep blue points show the mass function of the clusters identified by DBSCAN in the UMAP space of the $1\Mpch$ box. These points were clustered such that the number of clusters matched the number of halos identified by ROCKSTAR (case (i)). The green points show the mass function calculated for DBSCAN run on UMAP space such that the ellipse is identified as a single cluster (case (ii)). The pink points show the mass function of all the halos identified by ROCKSTAR within the $1\Mpch$ box.}
    \label{fig:halo_mass_func}
\end{figure}

\subsection{The Effect of $\nn$} \label{app:umap_params}
Figure \ref{fig:appendix_umap_nn} shows the change in the 2D reduced maps for varying values of $\nn$ at $z=0$. For the smallest value of $\nn$, apart from the four big structures (similar to the main text), the map is an almost uniform distribution of tiny structures comprising $\sim \nn$ particles. As the value of $\nn$ increases, the average size of clusters increases and are well separated from each other in the reduced maps. The change from $\nn = 4$ to $\nn = 20$ is significant, but from $\nn=20$ and $\nn=250$ is almost insignificant. By observing the largest cluster in each of the UMAP spaces, we see that for $\nn = 4$, the largest cluster corresponds to the central regions of a number of different infalling halos and subhalos. In contrast, the largest clusters in the other two maps correspond to the central region of the MW, in particular, the virialized region, as mentioned in the main text. Therefore, $\nn = 4$ maps find that all the cluster centers are similar and can be grouped as one. Depending on the problem one is trying to answer, different $\nn$ values could be useful.

From a visualization perspective, the large clusters approach a perfect circle at the $\nn$ value is increased. On a separate note, even though $\nn$ of 20 and 250 look similar in terms of the global patterns, they seem to be rotations of one another. This is attributed to the stochastic nature of the algorithm. One can negate the stochasticity at the cost of speed by foregoing multiple cores and setting a random seed.

\section{Cluster mass function} \label{app:mass_func}

Here, we discuss the cluster mass function. We perform DBSCAN clustering for two different sets of parameters - i) we adjust the number of clusters to match the number of halos identified by ROCKSTAR in the $1\Mpch$ box; in this case, $8.5\%$ of the particles are classified as noise, and ii) a set of parameters such that the entire large ellipse is classified as a single cluster, in this case, $10\%$ of particles are classified as noise. Note that in the discussions so far, we have been referring to case (ii) as the fiducial case. We then calculate the mass of the DBSCAN cluster as the sum of the masses of all the particles assigned to it. ROCKSTAR finds a total of 3711 halos inside our box. We find 3735 and 5647 in case (i) and case (ii), respectively. The mass function is shown in Figure \ref{fig:halo_mass_func}. The line corresponds to the analytical Tinker halo mass function. The pink points correspond to ROCKSTAR, the green points to UMAP clustering case (i), and the blue points to UMAP clustering case (ii). Both the ROCKSTAR and the UMAP halo mass functions lie above the Tinker theory curve. This is expected since we are looking at the mass function around the Milky Way, which is a biased region in the universe. However, we also find that the DBSCAN clusters on UMAP space, on average, appear more massive than the ROCKSTAR halos. It does, however, agree more closely with the mass function from ROCKSTAR based on the peak halo mass rather than $M_{\rm vir}$. The peak halo mass is the largest mass a halo in ROCKSTAR has ever attained over its history. 

\end{document}